\documentclass [twocolumn,showpacs,prl]{revtex4}
\usepackage{amssymb}
\usepackage{graphicx}
\usepackage{graphics}
\usepackage{bm}
\usepackage{verbatim}
\usepackage{epsf}
\begin{document}
\narrowtext
\title{ Sub-lambda gratings, surface plasmons, hotter electrons and brighter x-ray
 sources-  enhanced absorption of  intense, ultrashort laser light  by tiny
 surface modulations}
\author{Subhendu Kahaly \footnote{electronic address: skahaly@tifr.res.in} and G. Ravindra Kumar \footnote{electronic address: grk@tifr.res.in}}
\affiliation{Tata Institute of Fundamental Research, $1^{st}$ Homi Bhabha Road, Mumbai-400 005, India.}
\date{\today}
\newcommand{\e}{\mbox{\boldmath$\eta$}}
\newcommand{\x}{\mbox{\boldmath$x$}}
\newcommand{\si}{\mbox{\boldmath$\xi$}}

\begin{abstract}
We observe near 100 $\%$ absorption of light in intense ultrashort
laser plasma interaction in a metal coated (Au on glass)
sub-$\lambda$ grating structure under suitable conditions and the
subsequent 'hot' electron generation from the grating plasma. In the
low intensity regime we determine the conditions in which a
monochromatic infrared light ($\lambda$ = 800nm corresponding to the
central wavelength of the ultrashort laser that we used in
subsequent experiments) efficiently excites surface plasmon in the
grating. Then we study how the surface plasmon resonance condition
changes when we excite them using low intensity ultrashort pulses.
We look at the reflectivity of light varying the incident light
intensity over a wide range
($2\times{10^{12}}Wcm^{-2}$-$2\times{10^{15}}Wcm^{-2}$). The
reflectivity of grating with the resonance condition satisfied is
the lowest over the whole range of intensity. We compare the data
with those obtained from highly polished (~ $\lambda$/10) Au mirror
target under identical conditions. At high intensities we look at
the hard x-ray emission from both the targets with and without the
resonance condition. The hard X-ray spectrum shows a bimaxwellian
i.e two temperature hot electron distribution with a hotter
component present under the resonance condition while in all other
cases it shows the presence of only one low temperature component.
\end{abstract}
\maketitle

The interaction of intense, ultrashort laser pulses with optically
flat solid surfaces result in hot solid density plasmas. The
exotic state of hot, dense matter, thus created, is
hydrodynamically frozen in the time scale of the parent ultrashort
(subpicosecond) pulse rendering the plasma highly inhomogeneous.
These dense laser produced plasmas are sources of hot suprathermal
electrons and their signature bremsstrahlung hard x-ray radiation.
Understanding the generation of these hot electron pulses and
their transport through matter under extreme conditions together,
pose a difficult problem in itself which is of utmost importance
from the point of view of the proposed Fast Ignition scheme
\cite{Tabak} which holds the potential of realising laser fusion.
On the other hand, femtosecond x-ray pulse emission has recently
made it possible to monitor ultrafast motions in nature
\cite{Siders, Petruck}. Thus the need to explore novel ways to
optimize the hot electron generation as well as finding ways to
control it cannot be overemphasized. Ideally one would like to
couple all the incident light to the plasma, which generally is
not very keen on light absorption, and maximise the temperature
and number of hot electrons and consequently the hard x-ray
emission. In deuterium cluster gas jet almost 90$\%$ absorption of
laser energy has been reported \cite{Ditmire}. In solids attempts
on enhancing the continuum emission by structuring the target
surfaces are also reported. For example, recent literature reports
impressive enhancements in soft \cite{Murnane} and moderately hard
x-ray regions \cite{Wulker} using structured surfaces, viz.
gratings \cite{Murnane,Gauthier,Gordon}, ``velvet" coatings
\cite{Kulcsar}, porous and nanocylinder
\cite{Gordon,Nishikawa,Nishikawa2} targets. We have recently
observed that irradiating metal nanoparticles on a surface
tremendously enhances the hard x-ray bremsstrahlung from these
plasmas \cite{Rajeev}. Plasmas with varied scale-lengths are known
to yield significant enhancements at the cost of an increase in
the x-ray pulse duration \cite{Nakano, Bastiani}.

In this letter we report the highest absorption of incident laser
light on solid density plasma till date. Using a simplistic one
dimensional periodic system we experimentally pinpoint the window
where incident light is almost completely absorbed resulting in very
hot electrons at moderately low light intensity. We believe this is
the first systematic study of reflectivity of high intensity light
from sub-$\lambda$ gratings.

One of the intuitive ways to couple more light would be to
restrict light by some mechanism to localized regions on the
target, i.e to trap the incident electromagnetic radiation on the
target. Surface-bound electromagnetic charge density waves on
metals, so called surface plasmon polaritons (SPs), which when
excited lead to such localization. It needs a metal
{\boldmath($\epsilon_{m}=\epsilon^{'}_{m}+i\epsilon^{''}_{m}$,
$\epsilon^{'}_{m}<0$)} dielectric
{\boldmath($\epsilon_{d}=\epsilon^{'}_{d}+i\epsilon^{''}_{d}$,
$\epsilon^{'}_{d}>0$)} interface to excite a SP which is a
propagating solution of Maxwell equations that exponentially
decays in both directions away from the interface. The SP
dispersion curve,
{\boldmath$k_{SP}=k^{'}_{SP}+ik^{''}_{SP}=(\omega/c)\sqrt{\epsilon_{m}\epsilon_{d}/(\epsilon_{m}+\epsilon_{d})}$}
as is well known lies to the right of the light-line and thus for
the SP to interact with the incident EM radiation it needs a
supply of additional momentum. In our experiments we use a one
dimensional periodic sub-$\lambda$ gold grating to satisfy the
momentum matching condition. In case of a sinusoidal grating of
period d and amplitude (groove depth) small enough not to perturb
the SP dispersion relation we can write,
{\boldmath$k^{'}_{SP}=k_{0}\sin{\theta}+2\pi n/d$}, where
 {\boldmath $k_{0}=\omega/c$} is the wave vector of light incident at an angle
 {\boldmath$\theta$}. We look at reflectivity of light from the sub-$\lambda$
 grating under conditions that all the diffraction orders apart from the
specularly reflected one are quenched.
\begin{figure}[t]
\centering \vskip -.5cm\includegraphics [scale=0.3]{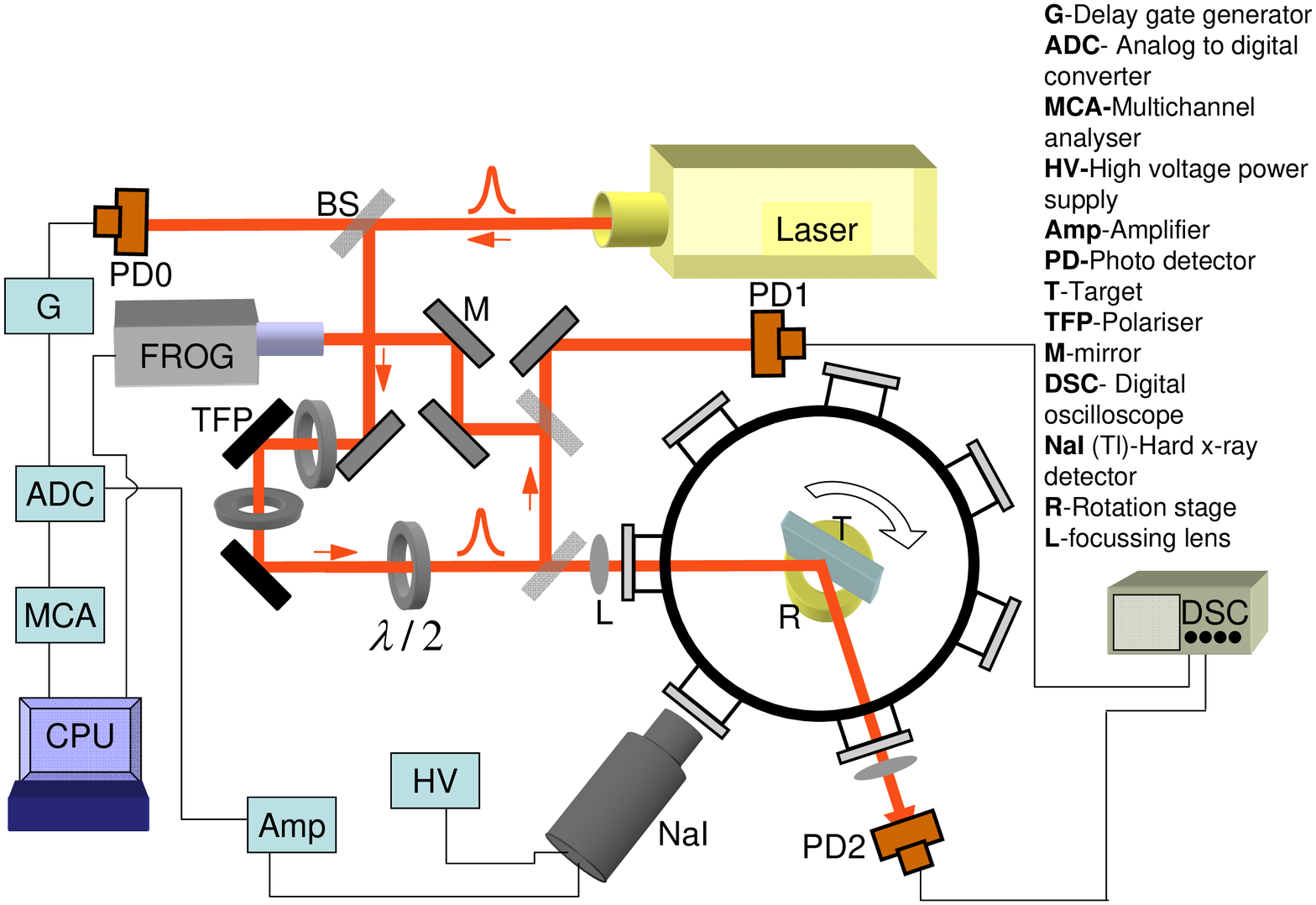}
\caption{Schematic of the experimental set up.}
\end{figure}

The experiments are performed using a Ti: Sapphire chirped pulse
amplified laser (Thales Laser, Alpha 10) emitting 55 fs pulses
centered at 800 nm wavelength at 10 Hz repetition rate. The laser
pulses have a contrast of 10$^{6}$ in picosecond timescales and
under optimum extraction, the prepulse (12 ns before the main
pulse) intensity level is less than 10$^{-6}$ of the main pulse.
The prepulse or the pedestal do not cause significant plasma
formation under our experimental conditions. The laser is focussed
at oblique incidence with a   f/20 lens on targets housed in a
vacuum chamber at 10$^{-4}$ torr. The target movement is
controlled  The maximum pulse energy used in the present set of
experiments, give a peak intensity of about 3 $\times$10$^{15}$
Wcm$^2$ at a 60 $\mu$m focal spot. The beam is intentionally
focussed loosely so as to have a precisely defined angle of
incidence by minimising beam divergence. A thin half wave plate in
the beam path selects the polarization state of light field. The
intensity of incident light is controlled by using high contrast
thin film polarizer and half wave plate combinations. The
experiments involving very low intensity laser pulses are
performed using unamplified nJ, 55fs modelocked pulse train (74
MHz) delivered by a Ti:sapphire oscillator. The target is
constantly translated in the focal plane in order to avoid
multiple laser hits at the same spot. In the series of low
intensity experiments ($<10^{9} Wcm^{-2}$) with continuous laser
or oscillator pulse train the reflected light is collected using a
calibrated power meter (Gentec). For measuring reflectivity as a
function of intensity, the reflected signal is collected using
integrating spheres and calibrated photodiodes. The hard x-ray
bremsstrahlung ( 25 keV - 250 KeV) emission under high intensity
laser irradiation is measured using a properly calibrated NaI(Tl)
scintillating detector looking along the plasma plume (i.e.
normally into the target). The data acquisition is time gated,
where the 30 $\mu$s collection time window is opened in
synchronization with the incident laser pulse. The hard x-ray
count rate is kept below one per second, i.e. 0.1 per laser shot
by adjusting the solid angle subtended into the detector in order
to avoid pile up problem in the detector.

In our experiments we used two types of targets. We used
triangular blazed gratings (period, \textbf{d} = 555 nm; groove
depth, \textbf{h} = 158nm, blaze angle 17.45$^{0}$) with gold
coating on glass substrate and compared all the data against
polished (roughness $\lambda$/10) gold coated glass targets.
\begin{figure}[t]
\centering
\includegraphics [scale=0.4]{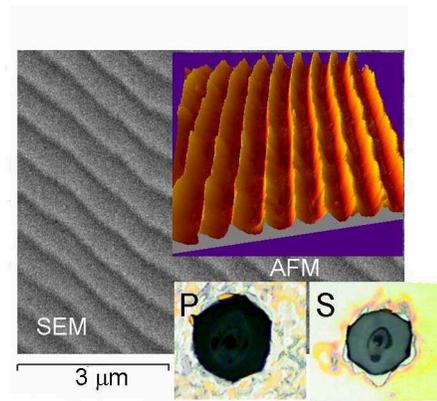} \includegraphics [scale=0.7]{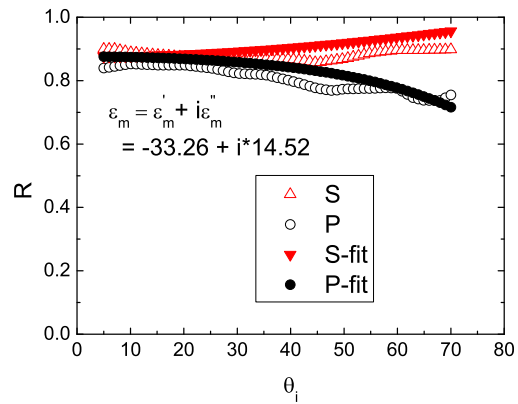}
\vskip -.5cm\caption{Above: SPM and AFM image of the sub-$\lambda$
grating. The lower inset shows grating image after laser
irradiation. Below: Experimental (hollow circles, triangles) and
calculated (solid symbols representing Fresnel fit) reflectivity
Vs. angle of incidence for flat Au mirror under low intensity
laser irradiation.}
\end{figure}
The thickness of gold layer in both the cases is many times greater
than the optical skin depth, {\boldmath$\delta_{s} \sim
c/\omega_{p}$} = 0.218 nm in our case ($\lambda$ = 800 nm).
Consequently, in all our experiments the glass background does not
have any significant role to play and the flat mirror as well as the
sub-$\lambda$ grating structure can be effectively assumed to be
made of gold only. The dielectric constant for Au ({\boldmath
$\epsilon_{m}$}) at $\lambda$ = 800 nm is extracted from the s and p
reflectivity vs. angle of incidence data (a typical one is shown in
Fig.2) by using Fresnel reflectivity formulae for an absorptive
medium with a complex refractive index \cite{BornWolf}. For Au metal
vacuum interface we find,
{\boldmath$\epsilon_{m}=\epsilon^{'}_{m}+i\epsilon^{''}_{m}$} =
-33.26 + \textbf{\emph{i}}14.52 which is reasonably close to the
values found in literature \cite{OSA}. The Drude model prediction,
{\boldmath$\epsilon^{'}_{m} = 1 - \omega_{p}^{2}/\omega^{2}$} =
-32.81  matches reasonably with the value we obtain experimentally
implying that at $\lambda$ = 800 nm there is no interband absorption
in metallic Au. In our experiments we kept the grating wave vector
aligned in the plane of incidence of light. A sub-$\lambda$
({\boldmath$\lambda/d >$} 1) grating permits only two diffraction
\begin{figure}[t]
\centering
\includegraphics [scale=0.8]{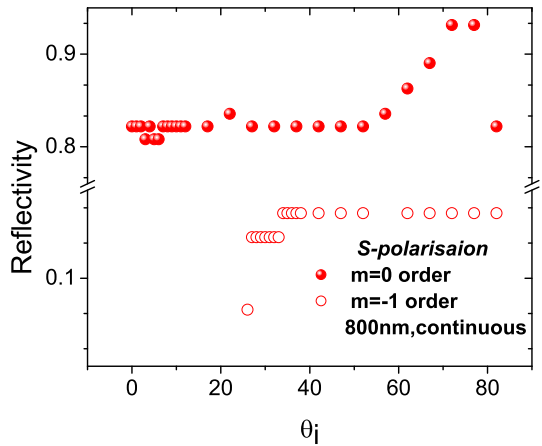}
\vskip -.5cm\caption{Diffraction efficiency (diffracted energy
normalized to incident energy) of the two available orders (m = 0
and m = -1) Vs. angle of incidence in case of the Au coated
sub-$\lambda$ grating structure under very low intensity
continuous s-polarised laser irradiation.}
\end{figure}
orders, i.e. the specularly reflected one m = 0 and the m =
-1$^{th}$ order \cite{Gauthier}. Fig.3 and Fig.4 shows the
fraction of incident energy diffracted by the grating into the two
available orders for s and p polarisations respectively as a
function of the angle of incidence. In the experiments with low
intensity CW light we note that for both the polarisations the
sub-$\lambda$ structure starts diffracting the -1$^{th}$ order
above a certain angle $\theta_{sp}$ = 23$\pm1^{0}$, below which we
only have specularly reflected light, i.e. below $\theta_{sp}$ the
grating ceases to be a diffracting medium and acts like a mirror
whose reflectivity is polarisation independent. For s
polarisation, light is mainly specularly reflected and the
reflectivity does not change with angle of incidence. In contrast,
the p polarised light shows a sharp dip (solid circles in Fig.4)
in specular reflectivity with the minimum at an angle which is
same as $\theta_{sp}$. At this angle almost all the light ($\sim
98.5 \%$) that is incident is completely absorbed. For p
polarisation $\theta_{sp}$ is the angle at which the phase
matching condition is satisfied and it corresponds to a situation
where the incident electromagnetic radiation efficiently
communicates with the surface charge density oscillations leading
to very strong excitation of surface plasmon, i.e. a condition
known as surface plasmon resonance(SPR). For continuous light, the
delta function like spectrum implies a very strict resonance
condition (phase matching condition) which manifests itself as a
very sharp dip in reflectivity around $\theta_{sp}$ (Fig.4: solid
circles). In the case of ultrashort pulse light ($\sim$ 55 fs),
the spectrum being broad ($\sim$ 50 nm bandwidth in our case) the
resonance condition is far more relaxed, and we observe a broader
SPR dip (Fig.4: crossed circles). The peak absorption is also
lower ($\sim 70 \%$) here owing to the fact that at SPR only the
central wavelength of the spectrum excites SP. The phase matching
condition for a low amplitude sinusoidal Au grating
\begin{figure}[t]
\centering
\includegraphics [scale=0.8]{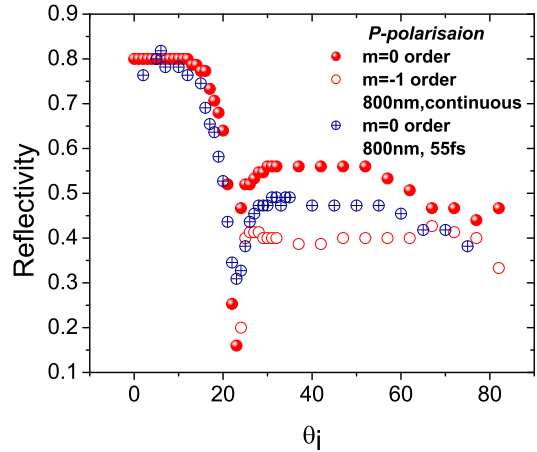}
\vskip -.5cm\caption{Diffraction efficiency (diffracted energy
normalized to incident energy) of various available orders Vs.
angle of incidence in case of the Au coated sub-$\lambda$ grating
structure under very low intensity continuous and 55 fs
p-polarised laser irradiation. Only two diffracted orders (m = 0
and m = -1) are seen with the dip corresponding to SP resonance
condition.}
\end{figure}
and vacuum interface is,
{\boldmath$k^{'}_{SP}=k_{0}\sin{\theta}+2\pi n/d$}, where
$k^{'}_{SP}=k_{0}.[(a+\sqrt{a^{2}+b^{2}})/2]^{1/2}$,
$a=[\epsilon^{'}_{m}(1+\epsilon^{'}_{m})+\epsilon^{''
2}_{m}]/[(1+\epsilon^{'}_{m})^{2}+\epsilon^{'' 2}_{m}]$ and
b=$\epsilon^{'}_{m}/[(1+\epsilon^{'}_{m})^{2}+\epsilon^{''
2}_{m}]$. With
{\boldmath$\epsilon_{m}=\epsilon^{'}_{m}+i\epsilon^{''}_{m}$} =
-33.26 + \textbf{\emph{i}}14.52, \textbf{d} = 555 nm and
{\boldmath$\lambda$} = 800 nm the SPR angle comes out to be
{\boldmath$\theta_{sp}$}=$26.56^{0}$ which is a close match to the
experimentally observed angle considering the fact that we have
used triangular gratings for our experiments. Fig.2 shows the SEM
and AFM image of the grating. We looked at the reflectivity of p
and s polarised light from both structued and polished flat
metallic surfaces at this particular angle of incidence
($23^{0}$), while varying the intensity of incident light over
three orders of magnitude (see Fig.5 and Fig.6). Flat Au mirror
reproduces the metallic reflectivity as measured in Fig2. for the
same angle of incidence and does so till the intensity reaches
\begin{figure}[t]
\centering
\includegraphics [scale=0.8]{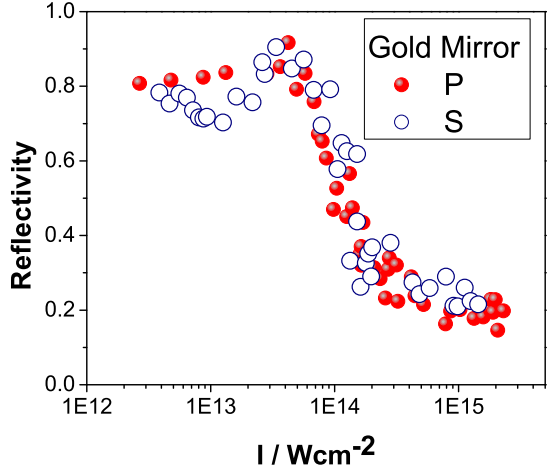}
\vskip -.5cm\caption{Gold reflectivity Vs. Intensity for S and P
polarisations. The knee corresponds to plasma formation threshold.}
\end{figure}
$I_{Th}\sim4.5\times10^{13} Wcm^{-2}$ which is the plasma
formation threshold for the material. Beyond this value plasma is
formed at the leading edge of the laser pulse and rest of the
light is reflected from the gold plasma which reflects less than
the gold metal. Under our experimental conditions the scale
length, $L = |n_{e}/\nabla n_{e}|\sim c_{s}\tau$ of the plasma is
very small and the p polarised light incident at a small angle of
$23^{0}$ does not create conditions favourable to resonance
absorption (RA) which excites the longitudinal electrostatic
charge density oscillations across the density gradient in the
highly inhomogeneous plasma \cite{Kruer} created by the ultrashort
pulse. Fig.5 shows that s and p reflectivity are the same within
the error bars of the experiment and corroborates this fact. As
shown in Fig.6 we observe that the grating reflectivity (Note that
only one diffraction order for this particular angle of incidence)
for s polarisation shows exactly the same behaviour as optically
flat Au with the same threshold for plasma formation in it. The
reflectivity of p polarised light from grating shows interesting
behaviour. At low intensity it starts from the value 0.33 which,
as expected, is roughly the same as the reflectivity of ultrashort
pulse assisting SPR as shown in Fig.4. The plasma formation
threshold in this case is dramatically reduced by a factor of 3 to
a value $I_{Th}\sim1.5\times10^{13} Wcm^{-2}$. Assuming that the
metallic sub-$\lambda$ structures under light irradiation, at
least till the moment plasma is formed, act as localized pockets
of enhanced electric fields, we expect f$\times I_{Th}$ to remain
the same for both the grating (G) as well as the mirror (M), i.e
$f_{G}.I_{Th,G}=f_{M}.I_{Th,M}$, where f is the absorption (1-R),
R being the reflectivity and $I_{Th}$ is the threshold of plasma
formation. In the case of p polarised light, $f_{G}$=0.67,
$f_{M}$=0.17 and $I_{Th,M}$=$4.5\times10^{13} Wcm^{-2}$ which
implies that, $I_{Th,G}$=$1.14\times10^{13} Wcm^{-2}$ matching
quite nicely with the experimentally observed value. The
absorption increases with increasing intensity and at about
$2\times10^{15} Wcm^{-2}$ reflectivity drops by 4.7 times its low
intensity value and almost 93$\%$ of the incident light is
absorbed. The absorbed light goes into the plasma and ultimately
manifests itself, among other things, as hot electrons. These hot
electrons leave the expanding plasma both towards the vacuum side
giving rise to ion acceleration due to capacitive effects, and
also towards the higher density solid region suffering collisions
with the charge centers and in the process emitting incoherent
bremsstrahlung hard x-ray and characteristic line radiations. The
bremsstrahlung photon energy distribution carries information
about the hot electron temperature and the bremsstrahlung yield
correspond to the number of hot electrons. Fig.7 shows the hard
x-ray spectrum obtained from high intensity (I = 3.8$\times10^{15}
Wcm^{-2}$) laser produced plasma on flat gold-coated glass target
and it shows the presence of only one temperature hot electrons,
though p-polarisation ($T_{h} =15\pm0.7$ KeV) produces hotter
electrons compared to s-polarisation ($T_{h}=11.8\pm0.4$ KeV).
\begin{figure}[t]
\centering
\includegraphics [scale=0.8]{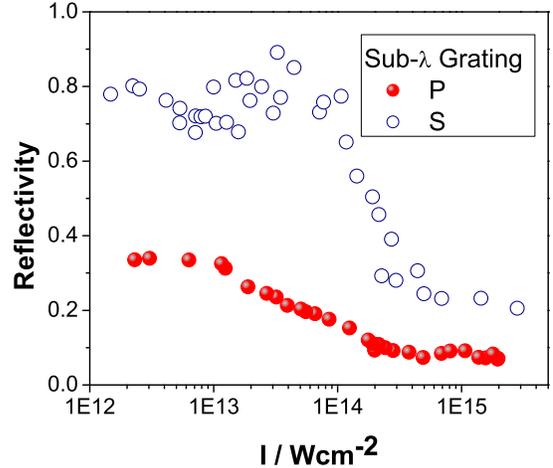}
\vskip -.5cm\caption{Grating reflectivity Vs. Intensity for s and
p polarisations. The knees correspond to plasma formation
thresholds in the two cases.}
\end{figure}
The lower half of Fig.7 shows hard x-ray emission spectrum from
the plasma generated on the grating surface when it is allowed to
interact with intense (I = 3.8$\times10^{15} Wcm^{-2}$) laser
incident at the SP resonance angle. s-polarisation gives rise to
maxwellian hot electron distribution with single temperature
component around $T_{h} = 16 \pm 1$ KeV while for p-polarisation
the spectrum is shows non-maxwellian behaviour with two
temperature hot electron distribution: one at $T_{h1} = 14 \pm 3$
KeV and the other at $T_{h2} = 67 \pm 11$ KeV the higher component
constituting 14$\%$ of the total number of hot electrons. The hard
x-ray yield for p-polarised light is 4.8 times that of s-polarised
case. That, the surface plasmon coupling plays the most crucial
role here (in the case of gratings) in the enhanced generation of
hotter electrons, is verified when we observe grating-plasma hard
x-ray spectrum ( at the same I = 3.8$\times10^{15} Wcm^{-2}$ )at
an angle of incidence that does not excite surface plasmon modes,
i.e. at an angle where the reflectivity dip is not there. The
Fig.9 shows one such spectrum at $\theta_{i} = 45^{0}$. Here,
s-polarisation produces one temperature electron distribution at
$T_{h} = 8.7 \pm 0.2$ KeV. p-polarisation gives rise to mainly
(99$\%$ weightage) one hot electron temperature at $T_{h} = 9 \pm
1$ KeV and a very feeble high temperature component around $26 \pm
11$ KeV.
\begin{figure}[t]
\centering
\includegraphics [scale=0.8]{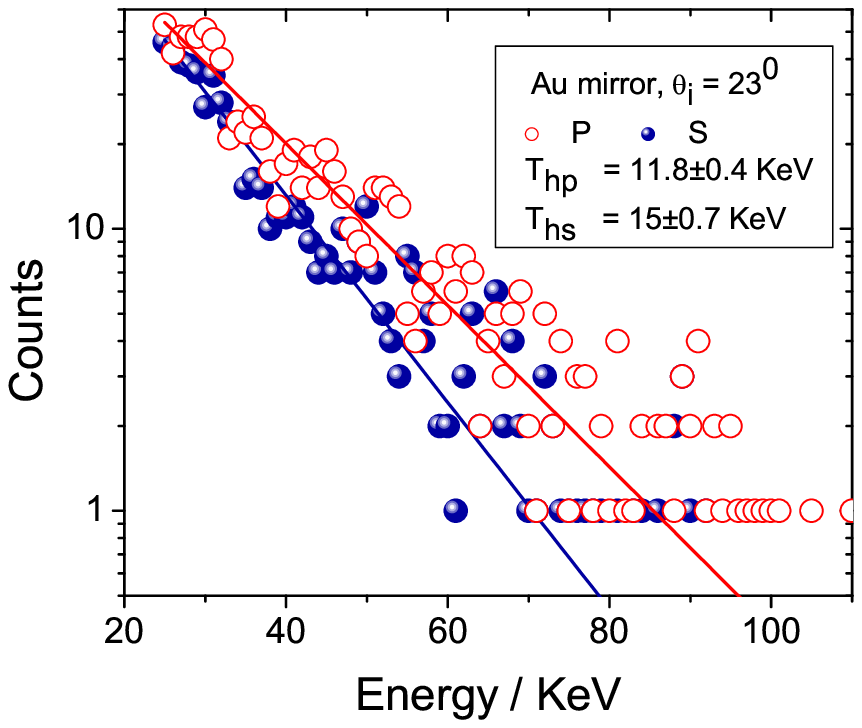} \includegraphics [scale=0.8]{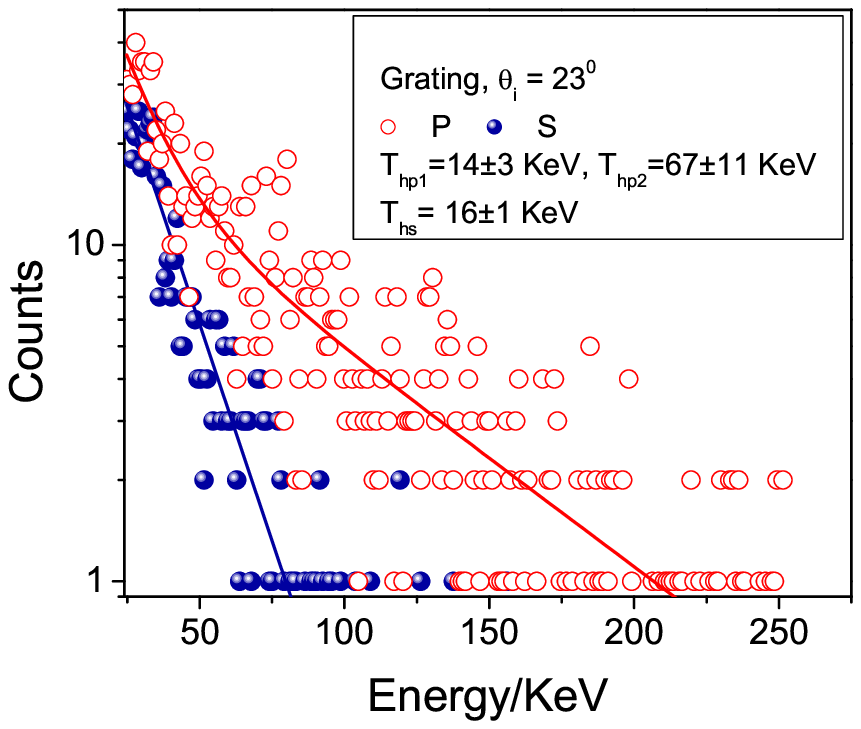}
\vskip -.5cm\caption{Bremsstrahlung emission spectra for Au mirror
plasma (Above) grating plasma (Below) with
$\theta_{i}=\theta_{sp}=23^{0}$. Solid curves denote least squares
temperature fits.}
\end{figure}

Thus we see that the surface plasmons that are efficiently excited
at resonance in the low intensity region maximise light absorption
in the structured metallic surface. In the low intensity regime
the excited nonradiative SPs relax by dumping their energy to the
background lattice and thus heating up the lattice. In our
experiments we see that it is possible to enhance to a great
extent the light coupling in the high intensity regime by
exploiting SPR. The very short length scale plasma retains the
imprint of grating structure in it's profile during its
interaction with the incident pulse. The reflectivity curves show
that the high level of light absorption is maintained even when we
increase the light intensity over 3 orders of magnitudes. The
lowering of the threshold of plasma formation in grating targets
under conditions favourable to SPR is indicative of the increase
in the effective field intensity due to the surface plasmon
assisted local field enhancement. Fig.2 inset shows the high
magnification optical microscopic image of the focal spot at the
sample after irradiation of the grating surface with high
intensity laser while maintaining the SPR angle, under two
polarisation conditions. In both the cases the magnification is
kept the same. Note that the pit formed due to p-polarisation
(SPR) is bigger and deeper compared to that due to s-polarisation
(No SPR) as expected from the reflectivity data. The grating
surface surrounding the focal spot is also much more damaged in p
irradiated spot, a feature that indicates violent excitation of
surface charge density oscillations when the grating is exposed to
incoming high intensity ultrashort laser pulse and larger length
of propagation of the generated SP wave along the metal vacuum
interface.
\begin{figure}[t]
\centering
\includegraphics[scale=0.8]{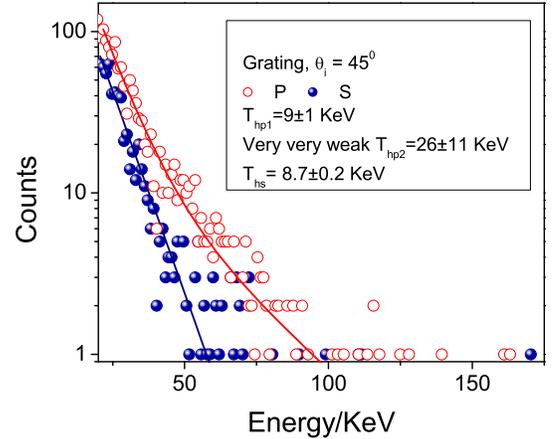}
\vskip -.5cm\caption{Bremsstrahlung emission spectra for grating
plasma with $\theta_{i}=45^{0}$. Solid curves are least squares
temperature fits.}
\end{figure}

In short, we have studied the reflectivity of incident light from
a 1-d sub-$\lambda$ periodic gold structure as a function of light
intensity, over three orders of magnitude, at an angle of
incidence that maintains SPR at low intensity. The light
absorption is compared to that of a flat Au target. We see almost
complete absorption of light. From the bremsstrahlung radiation at
high intensity we observe that SPR is assisting the generation of
hotter electrons in larger numbers. One can tune the hot electron
temperature by switching the polarisation state of the light under
SPR. We believe that this study is the first direct evidence that
SPs do couple the incident light and transfer the energy to the
plasma even at high intensities.

\end{document}